\documentclass[aps,prd,12pt,preprint,nofootinbib,superscriptaddress,titlepage,tightenlines,floatfix]{revtex4}

\usepackage{graphicx}
\usepackage{dcolumn}
\usepackage{bm}
\usepackage{epsfig}

\newcommand{\gev}{\,\mbox{GeV}}
\newcommand{\mev}{\,\mbox{MeV}}
\newcommand{\kev}{\,\mbox{keV}}
\newcommand{\nb}{\,\mbox{nb}}
\newcommand{\pb}{\,\mbox{pb}}
\newcommand{\invpb}{\,\mbox{pb}^{-1}}

\newcommand{\jpsi}{J/\psi}
\newcommand{\pp}{\psi(2S)}
\newcommand{\pdp}{\psi(3770)}

\newcommand{\pizero}{\pi^0}
\newcommand{\kstar}{{K^*}}

\newcommand{\ccbar}{c \bar c}
\newcommand{\ddbar}{D \bar D}

\newcommand{\ThreePi}{$\pi^+\pi^-\pizero $}

\newcommand{\RhoPiSum }{$\rho\pi $}
\newcommand{\RhoPiN}{$\rho^0 \pizero $}
\newcommand{\RhoPiC}{$\rho^+ \pi^- $}
\newcommand{\OmegaPiz}{$\omega\pizero $}
\newcommand{\PhiPiz}{$\phi\pizero $}
\newcommand{\RhoEta}{$\rho\eta $}
\newcommand{\OmegaEta}{$\omega\eta $}
\newcommand{\PhiEta}{$\phi\eta $}
\newcommand{\RhoEtaPrime}{$\rho\eta' $}
\newcommand{\OmegaEtaPrime}{$\omega\eta' $}
\newcommand{\PhiEtaPrime}{$\phi\eta' $}
\newcommand{\KstarKN}{$\kstar^0 \overline{K^0}$}
\newcommand{\KstarKC}{$\kstar^+ K^-$}
\newcommand{\KstarKSum}{$\kstar K$}
\newcommand{\BOnePiSum }{$b_1 \pi $}
\newcommand{\BOnePiN }{$b_1^0 \pizero $}
\newcommand{\BOnePiC }{$b_1^+ \pi^- $}

\begin{document}

\title{Decay of the $\pdp$ to Light Hadrons}

\author{G.~S.~Adams}
\author{M.~Anderson}
\author{J.~P.~Cummings}
\author{I.~Danko}
\author{J.~Napolitano}
\affiliation{Rensselaer Polytechnic Institute, Troy, New York 12180}
\author{Q.~He}
\author{H.~Muramatsu}
\author{C.~S.~Park}
\author{E.~H.~Thorndike}
\affiliation{University of Rochester, Rochester, New York 14627}
\author{T.~E.~Coan}
\author{Y.~S.~Gao}
\author{F.~Liu}
\affiliation{Southern Methodist University, Dallas, Texas 75275}
% ...

\collaboration{CLEO Collaboration} 
\noaffiliation

\date{September 9, 2005}

\begin{abstract} 

We describe a search for $\pdp$  
decay to two-body non-$\ddbar$ final states in
$e^+e^-$~data produced by the CESR
collider and analyzed with the CLEO-c detector. 
Vector-pseudoscalar production 
\RhoPiN, \RhoPiC, \OmegaPiz, \PhiPiz, 
\RhoEta, \OmegaEta, \PhiEta, 
\RhoEtaPrime, \OmegaEtaPrime, \PhiEtaPrime, 
\KstarKN, and \KstarKC\ is studied
along with that of \BOnePiSum\ (\BOnePiN\ and \BOnePiC) and \ThreePi. 
The largest amount of disagreement between the expected
rate for $e^+ e^- \to \gamma^* \to X$ 
and that for $e^+ e^- \to X$ at $\sqrt s = 3.773\gev$ 
is found for $X=\phi\eta$,
at an excess cross section of $(2.4 \pm 0.6)\,\mbox{pb}$ 
[$\Gamma_{\phi\eta}^{\pdp} =(7.4 \pm 1.6)\kev$], 
and a suggestive suppression is seen for \ThreePi\ and \RhoPiSum.
We conclude with form factor determinations  
for \OmegaPiz, \RhoEta, and \RhoEtaPrime. 
\end{abstract}

\vspace*{2mm}
\maketitle

The $\pdp$ charmonium state decays most copiously into the
OZI-allowed $\ddbar$~pair owing to the closeness of the mass threshold. 
Hadronic or radiative transitions to lower-lying charmonium states, 
decay to lepton pairs, or decay to light hadrons are all available,
but their branching fractions are highly suppressed. 
The $\pp$-$\pdp$ mixing scenario proposed in~\cite{mixing} gives rise to an
enhancement of certain partial decay widths, but 
the resulting branching fractions are still small
due to the large width of the $\pdp$. 
Nevertheless, some of the branching fractions are within experimental reach, 
and experimental information on $\pdp$ non-$\ddbar$ decays has 
recently begun to emerge~\cite{nonddbar}. 

This Letter describes the search for $\pdp$ decay to
vector pseudoscalar (VP) final states (\RhoPiN,
\RhoPiC, \OmegaPiz, \RhoEta, \OmegaEta, \PhiEta,
\RhoEtaPrime, \OmegaEtaPrime, \PhiEtaPrime, \KstarKN, \KstarKC) 
in CLEO-c data taken at the $\pdp$~resonance.  
We also seek \ThreePi\ as it is a mode that exhibits
curious structure in $\pp$~decay~\cite{cleovp}
and \BOnePiSum\ (in both the charged and the neutral
isospin submodes) 
as the most commonly produced
two-body hadronic final state in $\pp$~decays.
We use data samples taken at two energies,
$\sqrt s = 3.671$ and $\sqrt s = 3.773\gev$.
We establish event yields in both by counting events
satisfying the selection criteria detailed below, and subtracting
misreconstructed and therefore erroneously selected events. 
We measure the visible
cross section at both center-of-mass energies
for all modes. 
The sideband-subtracted event counts at $\sqrt s = 3.773\gev$ 
are compared with the expected rate from 
$e^+e^- \to \gamma^* \to \mbox{VP}$ (continuum), 
in order to discern a statistically significant discrepancy between the 
two.
Assuming the continuum cross section 
$\sigma (e^+e^- \to \gamma^* \to \mbox{VP})$ 
is given by 
\begin{equation}
\label{eqn:ff}
\sigma(s) = \frac{4\pi\alpha^2_{\mathrm{em}}}{3} 
            \frac {|{\cal F}(s)|^2 q_{\mathrm{VP}}^3(s)}{s^{3/2}},
\end{equation}
its measurement gives access to the form factor ${\cal F}$. 
The momentum of either hadron is denoted by $q_{\mathrm{VP}}$.
For all channels, the event yield at $\sqrt s = 3.671\gev$ is solely 
due to the above process. 
Also, for certain channels the event yield at $\sqrt s = 3.773\gev$ 
will be entirely attributable to continuum production, namely those that
cannot be produced through $\ccbar \to ggg$ because of isospin
suppression, such as \OmegaPiz, \RhoEta, and \RhoEtaPrime. 
Their remaining open avenue for $\pdp$~decay is 
$\ccbar \to \gamma^*$, which is severely suppressed. 

We use $e^+e^-$~collision data at
$\sqrt s = 3.773\gev$ (${\cal L} = 281\invpb$) and
$\sqrt s = 3.671\gev$ (${\cal L} = 21\invpb$).
The data analyzed here were
collected with the CLEO detector~\cite{cleo} operating at the Cornell
Electron Storage Ring (CESR)~\cite{cesr}.
The CLEO detector features a solid angle coverage of $93\%$ for
charged and neutral particles.
The charged particle tracking system operates in a 1.0~T~magnetic field
along the beam axis and achieves a momentum resolution of
$\sim 0.6\%$ at momenta of $1\gev/c$. The CsI crystal
calorimeter attains
photon energy resolutions of $2.2\%$ for $E_\gamma = 1\gev$
and $5\%$ at $100\mev$.
Two particle identification systems, one based on energy loss ($dE/dx$) in
the drift chamber and the other a ring-imaging Cherenkov (RICH)
detector, together are used to separate kaons from pions.
The combined $dE/dx$-RICH particle identification procedure has
a pion or kaon efficiency $>$90\% and a probability
of pions faking kaons (or vice versa) $<$5\%.

We identify intermediate states through the
following decays:
$\pi^0 \to \gamma\gamma$,
$\eta \to \gamma\gamma$ or $\pi^+\pi^-\pi^0$,
$\eta' \to \pi^+\pi^-\eta$ ($\eta \to \gamma\gamma$ only),
$\overline{K^0} \to  K_S \to \pi^+\pi^-$,
$\omega \to \pi^+\pi^-\pi^0$,
$\rho \to \pi^+\pi^-$,
$\phi \to K^+K^-$ or $\pi^+\pi^-\pi^0$,
$K^* \to K\pi$, and
$b_1 \to \omega\pi$.
Event selection proceeds exactly as for CLEO's
$\pp \to h_1 h_2$ analysis~\cite{cleovp}, the main features
of which were requirements on the total energy $E_{\mathrm{vis}}$ and 
momentum as well as on the invariant masses of intermediate particles,
in combination with particle identification criteria.
For some modes which are particularly
susceptible to background from radiative Bhabha or
$\mu^+\mu^-$~production, we tighten the selection
in the present analysis
by imposing the following additional requirements: 
For \RhoPiC\ and \ThreePi,
$e^+e^- \to \mu^+\mu^-(\gamma)$
events with a fake $\pizero$~candidate
are suppressed by a decay angle requirement 
of $|\cos\alpha|<0.8$.
For \RhoEta, backgrounds
from $e^+e^-(\gamma)$ final states with a fake
$\eta$~ candidate are reduced by allowing 
neither $\pi^\pm$ to satisfy electron identification
criteria.

\begin{figure}[h]
\includegraphics*[width=6.4in]{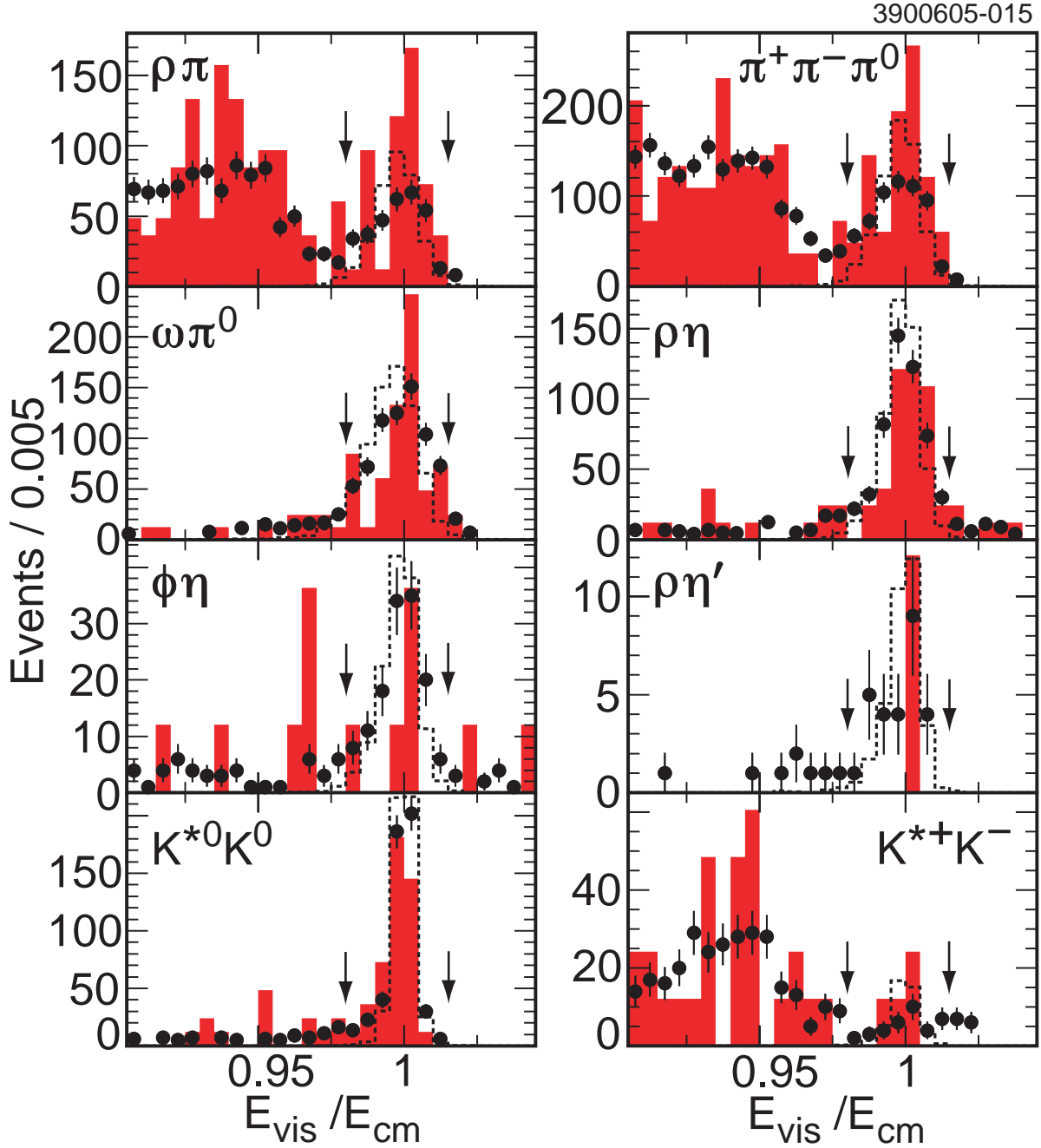}
\caption{Scaled visible energy $E_{\mathrm{vis}}/{\sqrt s}$ for selected
final states. Circles: data at $\sqrt s = 3.773\gev$, 
shaded histogram: data at $\sqrt s = 3.671\gev$ scaled by luminosity,
dashed histogram: signal~MC, arbitrary normalization. Arrows indicate
selection intervals.
\label{fig:xtot} }
\end{figure}

\begin{figure}[h]
\includegraphics*[width=6.4in]{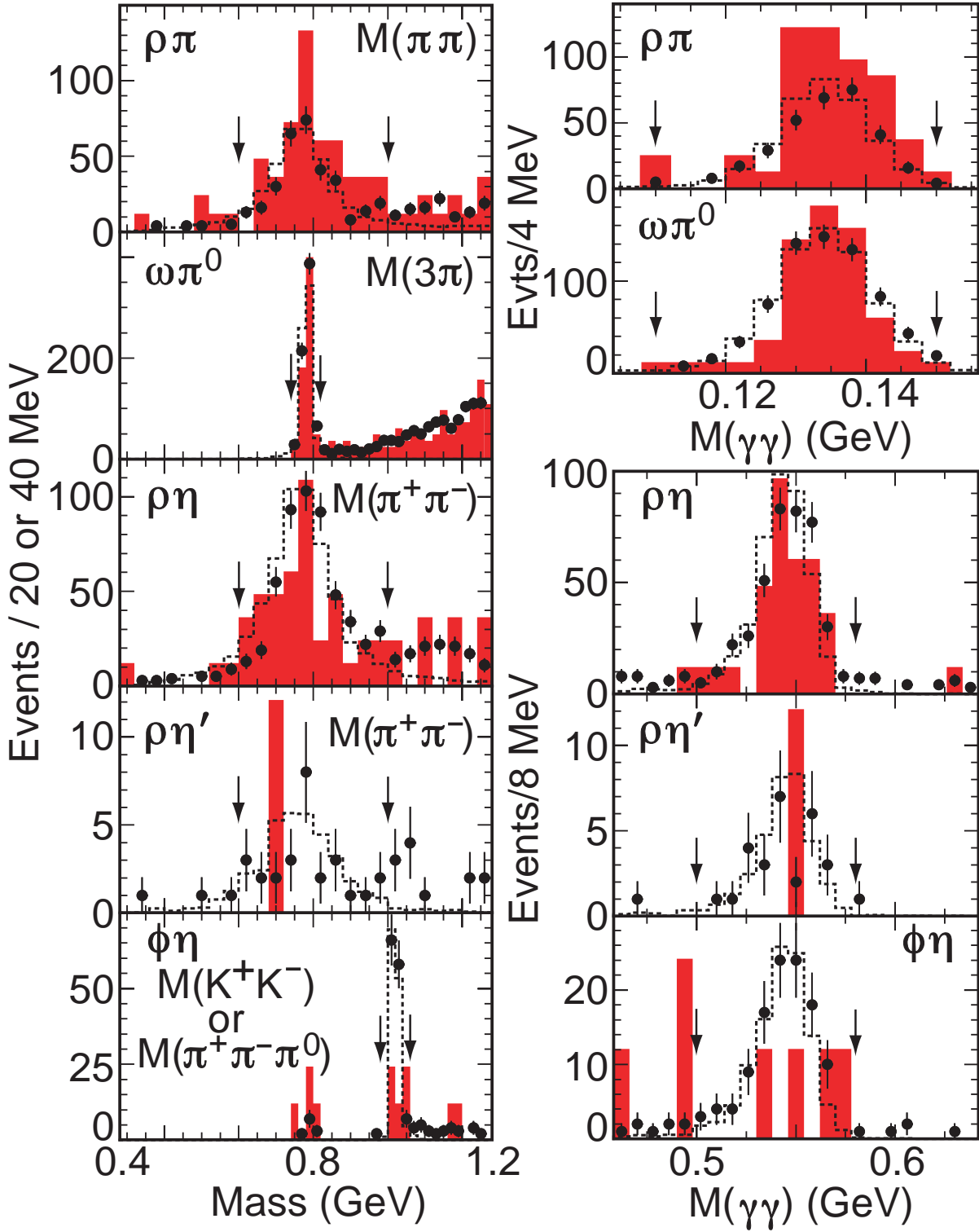}
\caption{Reconstructed invariant mass distributions for selected final states.
Symbols as in Fig.~1. The figures on the left (right)
side pertain to the first (second) of the two final state particles.
\label{fig:mx} }
\end{figure}

\begin{figure}[h]
\includegraphics*[width=6.4in]{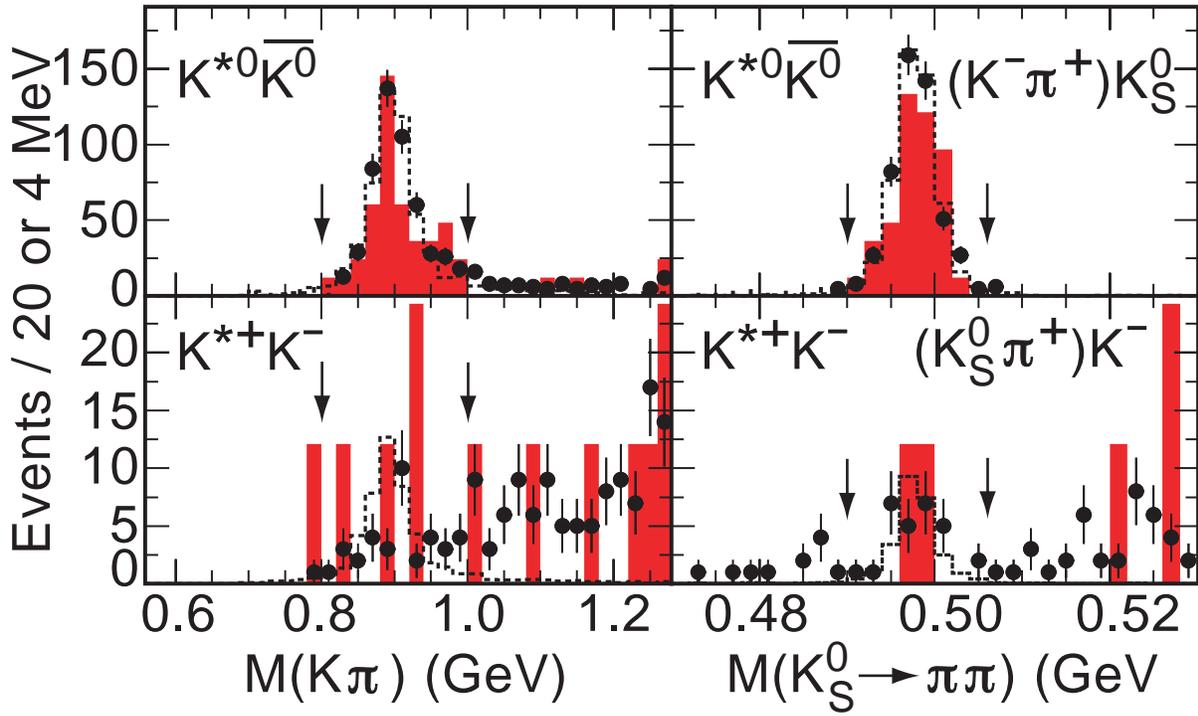}
\caption{Mass distributions for selected final states, continued.
Symbols as in Fig.~1.
\label{fig:mxkaons} }
\end{figure}

We present distributions of scaled total energy $E_{\mathrm{vis}}/{\sqrt s}$ 
and reconstructed invariant masses for selected modes
in Figures~\ref{fig:xtot}-\ref{fig:mxkaons}. All cuts other than that 
imposed on the quantity displayed have been applied.

The efficiency $\epsilon$ for each final state is obtained from 
signal Monte Carlo with the {\tt EvtGen}~event generator~\cite{EvtGen},
including final state radiation~\cite{PHOTOS}, and a {\tt GEANT}-based 
detector simulation~\cite{GEANT}. 
We generated the VP~modes with
angular distribution $(1+\cos^2\theta)$~\cite{BRODLEP}, $b_1\pi$ flat
in $\cos\theta$, and $\pi^+\pi^-\pi^0$
as in $\omega$ decay. We assume ${\cal B}(b_1\to\omega\pi)$=100\%.

Systematic uncertainties on the cross section measurements
arise from various sources, some
common to all channels, some channel specific:
The systematic errors on branching fraction ratios
share common contributions from the uncertainty
in luminosity (1\%), 
trigger efficiency (1\%), 
and electron veto (0.5\%).
Other sources vary by channel, including cross-feed adjustments
(50\% of each subtraction), MC statistics, 
accuracy of MC-generated polar angle and mass distributions
(5\% for $b_1\pi$, 14\% for $\pi^+\pi^-\pi^0$),
and detector performance modeling quality: charged particle
tracking (1\%/track),
$\pi^0$/$\eta$ and $K_S$ finding (2\%/($\pi^0$/$\eta$), 5\%/$K_S$),
$\pi/K$ identification (3\%/identified $\pi/K$), and resolutions of
mass (2\%) and total energy (1\%).

Systematic uncertainties dominate in the cross section measurements
for most channels at $\sqrt s = 3.773\gev$~data and are comparable to 
the statistical errors for some modes at $\sqrt s = 3.671\gev$. 
 
The signal yields at both center-of-mass energies are listed
in Table~\ref{tab:evtyields}, separated into signal mass
windows and sideband counts. Also listed are the efficiencies
and cross sections. The statistical errors arise from 
68\%~CL~intervals. All cross sections include an upward correction
of $(20 \pm 7)$\%~to account for initial and final state radiation 
effects~\cite{cleovp}.
In case of the isospin-violating modes, we also correct for 
electromagnetic interference
between the tails of the $\jpsi$, $\pp$, and $\pdp$ resonances with
continuum production by a 4.9\% upward [1.2\% downward]
adjustment to the cross-sections at $\sqrt s = 3.671\gev$ 
[$\sqrt s = 3.773\gev$]~\cite{interferencecorrection}. 
The results in Table~\ref{tab:evtyields} for $\sqrt s = 3.671\gev$
supersede those in~\cite{cleovp}.

We now focus on the discrepancy between the $\sqrt s = 3.773\gev$ 
yield and expected continuum contribution in order to determine
whether there is significant production from $\pdp$~decays.
To arrive at an estimate for the continuum background at
$\sqrt s = 3.773\gev$, two routes are pursued:
\underline{Method~I.}~We scale the measured yield 
(after sideband subtraction) at $\sqrt s = 3.671\gev$ 
by the luminosity ratio, the ratio of efficiencies ($0.88-1.00$),
and an assumed dependence of $1/s^3$ of the continuum
cross section, corresponding to a form factor dependence
of $1/s$. This method uses data as much as possible, 
but suffers from the limiting event yield in the lower energy 
data sample. Using a different power $1/s^n$ results
in a relative change of 5.4\% in the scale factor per unit of~$n$. 
\underline{Method~II:}~We 
use a SU(3)-based scaling prediction, whereby the
the cross sections $\sigma( e^+e^- \to \mathrm{VP} )$
are linked~\cite{RATIOS} as
$\omega\pi : \rho\eta : K^{*0}\bar{K^0} 
: \rho\pi : \rho\eta' : \phi\eta 
: K^{*+}K^- : \phi\eta' : \omega\eta 
: \omega\eta' : \phi\pi
= 
1:2/3:4/9
:1/3:1/3:4/27
:1/9 :2/27:2/27
:1/27:0$.
By combining our data of the two isospin violating modes with 
highest statistics, \OmegaPiz\ and \RhoEta\ (scaled up by a
factor of 3/2),
we determine a unit of cross section as 
$\sigma^{SU(3)}= (15.1 \pm 0.5) \pb$ at $\sqrt s = 3.773\gev$.
This results in a precise prediction for each channel,
albeit a model-dependent one. 
We note satisfactory agreement between the yields expected on this
basis and those observed in the data at $\sqrt s = 3.671\gev$ for all
channels except of \KstarKSum\ (also see~\cite{cleovp}).
No such prediction is made for \ThreePi\ and~\BOnePiSum. 
 
For each channel, both continuum predictions are compared with 
the yield at $\sqrt s = 3.773\gev$ by a method similar to that proposed 
in~\cite{FELDCOUS}. It is the same procedure that was applied
to $\pp$~decays in~\cite{cleovp}: 
The probability that the
the continuum production from either method together with the
misreconstruction background as estimated from the sidebands
fluctuate to an event count equal to or beyond the observed signal yield
at $\sqrt s = 3.773\gev$ 
is calculated with simulated trials 
governed by Poisson statistics. 
These, expressed in units of standard deviations, are included as
$S^{\mathrm{I}}$ and $S^{\mathrm{II}}$ in Table~\ref{tab:evtyields}.
We find statistical agreement between the yields, with a few exceptions.
The mode \PhiEta\ is found to be enhanced over the prediction 
from either method:
The weighted mean excess over continuum production
is $(61.6 \pm 11.7)$ events. This corresponds to
a cross section of $\sigma^{\pdp}_{\phi\eta}=
(2.4 \pm 0.5 \pm 0.3)\pb$, or, using
$\sigma^{\mathrm{obs}}(\pdp \to \ddbar) 
= (6.39 \pm 0.20) \nb$~\cite{sigmaddbarcleo}
and removing the radiative correction factor in $\sigma^{\pdp}_{\phi\eta}$, 
a branching fraction ${\cal B}(\pdp \to \phi\eta) = 
( 3.1 \pm 0.6 \pm 0.3  \pm 0.1 )\times 10^{-4}$, where the
first error is statistical, the second systematic arising 
from this measurement, and the third that induced by 
$\sigma^{\pdp}_{\ddbar}$.
A partial width of $\Gamma_{\phi\eta}^{\pdp} = (7.4 \pm 1.6)\kev$
follows~\cite{pdg2004}.
Suggestive suppressions are observed for \ThreePi and \RhoPiN. 
The observed \KstarKN\ cross section at $\sqrt s=3.773\gev$ is
consistent
with being saturated by the expected continuum production
as extrapolated from $\sqrt s=3.671\gev$ (Method~I).
As the observed \KstarKN\ cross sections at {\it both}
energies~\cite{cleovp} far exceed the SU(3) predictions (Method~II),
and by similar
amounts, there is no indication of a substantial
$\pdp \to$\KstarKN\ contribution.
Rather, the excesses originate in the continuum process,
$e^+e^- \to$\KstarKN.
The same comments apply to \KstarKC, but with respect to an
observed deficit obtained with Method~II.
 
\begin{figure}[h]
\includegraphics*[width=6.4in]{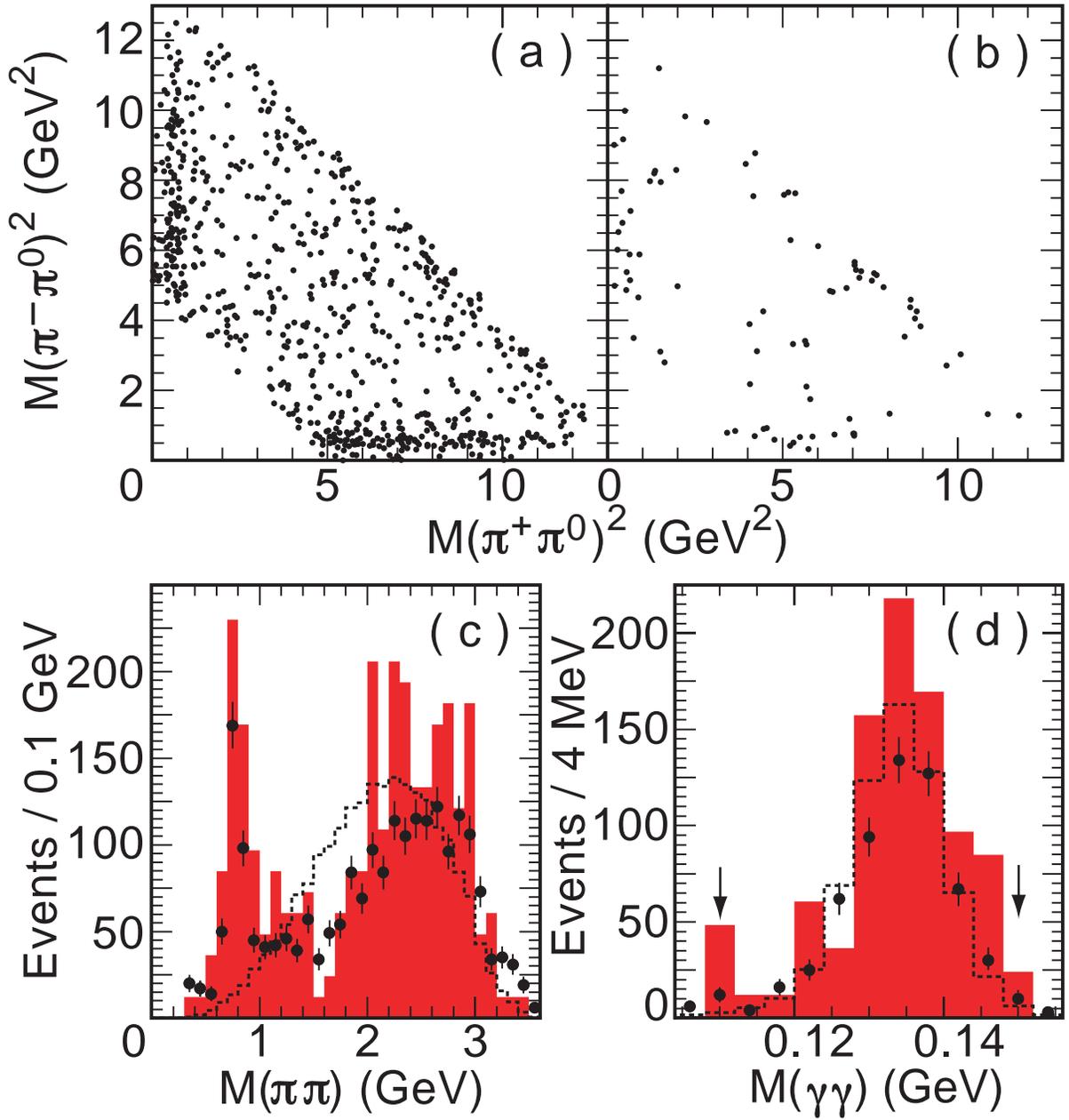}
\caption{Dipion invariant mass distributions for the \ThreePi\ final state
in data (a) at $\sqrt s = 3.773\gev$, (b) at $\sqrt s = 3.671\gev$. 
(c) The invariant mass of all pion
pairs per event and (d) the reconstructed $\pi^0$~mass, in 
data at $\sqrt s = 3.773\gev$ (circles), 
data at $\sqrt s = 3.671\gev$, scaled by luminosity (shaded histogram), 
and phase space MC (dashed line). 
\label{fig:threepi} }
\end{figure}

Additional information on \ThreePi\ is shown in Figure~\ref{fig:threepi}.
The dipion invariant masses in $\sqrt s = 3.773\gev$~data shows features 
similar to that of $\sqrt s = 3.671\gev$
({\sl i.e.,} population of the $\rho$~mass
bands together with an accumulation at higher masses); 
the yield reduction appears uniform in the dipion invariant mass
distribution. 

We compute upper limits on the event yields originating 
from $\pdp$~decays for all modes,
where we treat those with a deficit as zero counts, neglecting
interference effects, and arrive at upper limits on the
observable cross section excess over continuum as included
in Table~\ref{tab:evtyields}. 

The measured cross sections for \OmegaPiz, \RhoEta,
and \RhoEtaPrime\ are converted into form factor measurements,
which are listed in Table~\ref{tab:ff}. Our results 
are in agreement with, but more precise than, those recently
reported by BES~\cite{besiv}.

In summary, we have sought twelve vector pseudoscalar final
states in data at $\sqrt s = 3.773\gev$. 
Combined with data collected at $\sqrt s = 3.671\gev$, we
establish cross section measurements for these channels
at both energies. We find evidence for the decay
$\pdp \to \phi\eta$, and see hints that $\pdp$~decays
to \RhoPiSum\ and \ThreePi\ could be causing a deficit to
appear in their yields through negative interference
with continuum production~\cite{mixing,wmyrhopi}. 
Otherwise, we note broad agreement with the continuum
predictions. 
Form factor measurements for \OmegaPiz, \RhoEta,
and \RhoEtaPrime\ have been presented. 
All our measurements are either firsts of their kind or constitute
an improvement over previous measurements.

\pagebreak

\begin{acknowledgments}
We gratefully acknowledge the effort of the CESR staff
in providing us with
excellent luminosity and running conditions.
This work was supported by
the National Science Foundation,
the U.S. Department of Energy,
the Research Corporation,
and the
Texas~Advanced~Research~Program.

\end{acknowledgments}

\begin{table*}
\centering
\caption{The number of events $N$ in the mass signal windows (``sw'')
and sidebands (``sb'') in data taken at $\sqrt s = 3.671\gev$ and 
$\sqrt s = 3.773\gev$ data; 
the efficiency $\epsilon$ in percent; 
the level of consistency or significance, expressed in units
of standard deviations, between continuum background and
observed yield, for the two methods of determining the
continuum background described in the text, $S^{\mathrm{I}}$
and $S^{\mathrm{II}}$;
the cross sections at $\sqrt s = 3.671\gev$ and 
$\sqrt s = 3.773\gev$;
the cross section $\pdp \to h_1 h_2$, computed as
the excess over the continuum prediction
as established using Method~I or Method~II (see text).
\label{tab:evtyields}
}
\footnotesize
\begin{tabular}{l|cc|cc|c|cc|cc|cc}
Channel 
        & $N^{3.67}_{\mathrm{sw}}$ & $N^{3.67}_{\mathrm{sb}}$
        & $N^{3.77}_{\mathrm{sw}}$ & $N^{3.77}_{\mathrm{sb}}$ 
        & $\epsilon$ 
        & $S^{\mathrm{I}}$ & $S^{\mathrm{II}}$ 
        & $\sigma^{\mathrm{3.67GeV}}\ [\pb]$ 
        & $\sigma^{\mathrm{3.77GeV}}\ [\pb]$ 
        & $\sigma_{\pdp}^{\mathrm{I}}\ [\pb]$ 
        & $\sigma_{\pdp}^{\mathrm{II}}\ [\pb]$ \\
\hline
      \rule[0.5mm]{0mm}{3.0mm}

        \ThreePi &       74 &      6.8
 &      576 &     72.3
 &     29.0
 &     -2.7
 &     
 & $    13.1^{+   1.9}_{-1.7}  \pm    2.1 $ 
 & $     7.4 \pm    0.4  \pm    1.2 $ 
 & $<$     0.04 & 
\\ 
       \RhoPiSum &       43 &      5.4
 &      314 &     44.8
 &     26.3
 &     -2.2
 &     -1.7
 & $     8.0^{+   1.7}_{-1.4}  \pm    0.9 $ 
 & $     4.4 \pm    0.3  \pm    0.5 $ 
 & $<$     0.04 & $<$    0.04 \\ 
   \quad\RhoPiN &       21 &      3.4
 &      130 &     33.0
 &     32.5
 &     -2.2
 &     -2.1
 & $     3.1^{+   1.0}_{-0.8}  \pm    0.4 $ 
 & $     1.3 \pm    0.2  \pm    0.2 $ 
 & $<$     0.03 & $<$    0.03 \\ 
    \quad\RhoPiC &       22 &      2.0
 &      184 &     11.8
 &     23.1
 &     -0.9
 &     -0.5
 & $     4.8^{+   1.5}_{-1.2}  \pm    0.5 $ 
 & $     3.2 \pm    0.3  \pm    0.3 $ 
 & $<$     0.05 & $<$    0.05 \\ 
       \OmegaPiz &       54 &      6.2
 &      696 &     39.2
 &     19.0
 &      0.9
 &     -0.2
 & $    15.2^{+   2.8}_{-2.4}  \pm    1.5 $ 
 & $    14.6 \pm    0.6  \pm    1.5 $ 
 & $<$      4.5 & $<$    0.06 \\ 
         \PhiPiz &        1 &      1.6
 &        2 &      4.0
 &     16.5
 &      0.0
 &     -0.0
 & $ <      2.2 $ 
 & $ <      0.2 $ 
 & $<$      0.2 & $<$     0.2 \\ 
         \RhoEta &       36 &      3.1
 &      508 &     31.0
 &     19.6
 &      1.1
 &      0.7
 & $    10.0^{+   2.2}_{-1.9}  \pm    1.0 $ 
 & $    10.3 \pm    0.5  \pm    1.0 $ 
 & $<$      4.0 & $<$     1.3 \\ 
       \OmegaEta &        4 &      0.0
 &       15 &      6.0
 &      9.9
 &     -1.7
 &     -2.9
 & $     2.3^{+   1.8}_{-1.0}  \pm    0.5 $ 
 & $     0.4   \pm 0.2 \pm    0.1 $ 
 & $<$      0.1 & $<$     0.1 \\ 
         \PhiEta &        5 &      1.0
 &      132 &     15.9
 &     11.0
 &      2.5
 &  $\geq 5 $ 

 & $     2.1^{+   1.9}_{-1.2}  \pm    0.2 $ 
 & $     4.5 \pm    0.5  \pm    0.5 $ 
 & $<$      4.5 & $<$     3.3 \\ 
    \RhoEtaPrime &        1 &      0.0
 &       27 &      0.9
 &      2.9
 &      1.0
 &     -1.3
 & $     2.1^{+   4.7}_{-1.6}  \pm    0.2 $ 
 & $     3.8^{+   0.9}_{-0.8}  \pm    0.4 $ 
 & $<$      4.7 & $<$     0.4 \\ 
  \OmegaEtaPrime &        0 &      0.0
 &        2 &      0.0
 &      1.5
 &  $\geq 5 $ 
 &      0.0
 & $ <     17.1 $
 & $     0.6^{+   0.8}_{-0.3}  \pm    0.6 $ 
 & $<$      3.0 & $<$     1.9 \\ 
    \PhiEtaPrime &        0 &      0.0
 &        9 &      2.0
 &      1.2
 &      2.4
 &      1.2
 & $ <     12.6 $
 & $     2.5^{+   1.5}_{-1.1}  \pm    0.4 $ 
 & $<$      5.2 & $<$     3.8 \\ 
        \KstarKN &       38 &      0.4
 &      501 &     18.1
 &      8.8
 &      1.1
 &  $\geq 5 $ 

 & $    23.5^{+   4.6}_{-3.9}  \pm    3.1 $ 
 & $    23.5 \pm    1.1  \pm    3.1 $ 
 & $<$      9.0 & $<$    20.8 \\ 
        \KstarKC &        4 &      1.0
 &       36 &     32.4
 &     16.0
 &     -1.4
 &     -4.1
 & $     1.0^{+   1.1}_{-0.7}  \pm    0.5 $ 
 & $ <      0.6 $
 & $<$      0.1 & $<$    0.1 \\ 
      \BOnePiSum &       20 &      4.5
 &      268 &    100.3
 &     11.3
 &     -0.1
 & 
 & $     7.9^{+   3.1}_{-2.5}  \pm    1.8 $ 
 & $     6.3 \pm    0.7  \pm    1.5 $ 
 & $<$      0.1 &
\\ 
   \quad\BOnePiN &        5 &      3.0
 &       49 &     82.5
 &      4.2
 &     -1.2
 &
 & $ <     17.1 $ 
 & $ <      2.5 $ 
 & $<$      0.4 & 
\\ 
   \quad\BOnePiC &       15 &      1.5
 &      219 &     17.8
 &     18.4
 &      1.0
 &  
 & $     4.2^{+   1.6}_{-1.2}  \pm    0.6 $ 
 & $     4.7 \pm    0.4  \pm    0.6 $ 
 & $<$      2.7 & 
\\ 

\end{tabular}
\end{table*}

\begin{table}
\centering
\caption{Form factors with statistical and systematic errors.}
\vskip 0.1 in
\begin{tabular}{l|c|c} 
         Channel & \multicolumn{2}{c}{${\cal F}(s)\ (\mbox{TeV}^{-1})$} \\
                 & $\sqrt s = 3.671\gev$ 
                 & $\sqrt s = 3.773\gev$  \\
 \hline 
       \OmegaPiz &  $ 40^{+4}_{-3} \pm 2 $ \rule[0.5mm]{0mm}{3.0mm}
                 &  $ 39 \pm 1         \pm 2 $  \\ 
         \RhoEta &  $ 34^{+4}_{-3} \pm 2 $ 
                 &  $ 34 \pm 1         \pm 2 $ \\
    \RhoEtaPrime &  $ 17^{+14}_{-9} \pm 1 $
                 &  $ 22^{+3}_{-2} \pm 1 $  \\
\end{tabular} 

\label{tab:ff}
\end{table}
 
\end{document}